\documentstyle[pra,aps]{revtex}

\begin{document}

\def\beqn{\begin{equation}}
\def\eeqn{\end{equation}}
\def\beqa{\begin{eqnarray}}
\def\eeqa{\end{eqnarray}}

\title{Quantum Langevin equations and stability }
\author{Marc-Thierry Jaekel $^a$ and Serge Reynaud $^b$}
\address{$(a)$ Laboratoire de Physique Th\'eorique de l'ENS \thanks{%
Unit\'e propre du CNRS associ\'ee \`a l'Ecole Normale Sup\'erieure et \`a %
l'Universit\'e Paris-Sud}, 24 rue Lhomond F75231 Paris Cedex 05 France\\
$(b)$ Laboratoire de Spectroscopie Hertzienne de l'ENS \thanks{%
Unit\'e de l'Ecole Normale Sup\'erieure et de l'Universit\'e Pierre et Marie
Curie associ\'ee du CNRS }, 4 place Jussieu F75252 Paris Cedex 05 France}
\date{{\sc Journal de Physique I} {\bf 3} (1993) 339-352}
\maketitle

\begin{abstract}
Different quantum Langevin equations obtained by coupling a particle
to a field are examined.
Instabilities or violations of causality affect the motion of a
point charge linearly coupled to the electromagnetic field.
In contrast, coupling a scatterer with a reflection cut-off to
radiation pressure leads to stable and causal motions. The radiative
reaction force exerted on a scatterer, and hence its quasistatic
mass,  depend on the field state.
Explicit expressions for a particle scattering a thermal field in a
two dimensional space-time are given.

\end{abstract}

\def\dq{\dot q}
\def\dphi{\dot \phi}
\def\dpsi{\dot \psi}
\def\ddq{\ddot{q}}
\def\dddq{\stackrel{...}{q}}
\def\dA{\dot A}

\section{INTRODUCTION}

Since its introduction for a description of Brownian motion
\cite{Langevin 1908}, Langevin
equation
 has been extended to a large variety of
domains \cite{lc} and has led to many mathematical developments \cite
{Arnold}.

In a general way, Langevin equation  describes the motion of
a small system, with a few degrees of freedom (for instance its
position ${\bf q}$), interacting with
a bath
composed of a very large number of degrees of freedom. Hence, the
bath can be considered to exert a fluctuating force on the small
system (of mass $m$). The force depends on the motion of the system,
 and for small displacements can be developed as
the fluctuating force experienced by the system at rest
(${\bf F}$),
plus a motional force proportional to the system's displacement:
\beqn
\label{le}
m\ddot{{\bf q}}(t) = {\bf F}(t) + \int_{-\infty}^\infty dt' \chi (t-t') {\bf
q}(t')
\eeqn
The fluctuating force (${\bf F}$) is characterised by its time
correlations, which are related to the motional susceptibility
($\chi$) through a fluctuation- dissipation relation. For  Brownian motion, the
force fluctuations have
 white noise correlations, and are  linked to the
frictional force, which is  proportional to the particle's velocity
($\dot{\bf q}$) \cite{Einstein,lc}.
Quantum versions have been developed, which preserve the main
features of Langevin equation \cite{lq}.

A charge coupled to a fluctuating electromagnetic field provides a
natural example of a system obeying Langevin equation \cite{EH}.
In this case, the frictional force is proportional to the third time
derivative of the charge's position ($\stackrel{...}{\bf q}$). As is
well known from classical electron theory, this
reaction force is plagued with instabilities. The equations of motion
possess `runaway solutions', i.e. exponentially self accelerating
motions. If specific boundary conditions are imposed to forbid such
unstable solutions, pre-acceleration effects occur, the particle's
motion anticipating on the applied force \cite{Rohrlich}.

Even in vacuum and for neutral bodies, field fluctuations lead to
macroscopic effects. Casimir forces and vacuum friction are such
manifestations due to radiation pressure fluctuations \cite{Casimir,de
Witt}. Objects which scatter a quantum field in vacuum experience a
frictional force when moving with non uniform acceleration \cite{FD}.
A perfect reflector for a scalar field in a two dimensional (2d)
space-time is submitted to a force proportional to the third time
derivative of its position ($\stackrel{...}{\bf q}$).
This force can be understood as the cumulative effect of radiation
pressure fluctuations, and satisfies a fluctuation- dissipation
relation
\cite{JR3}.

The introduction of a frequency dependent scattering, with causality,
unitarity and a high frequency transparency conditions, has provided
a simple remedy for divergences induced by the vacuum fluctuations, of
infinite energy \cite{JR2}.  This description also gives a
treatment of Langevin equation related with vacuum
radiation pressure, which is consistent and free from instabilities \cite{JR5}.

This approach is applied here to Langevin equations derived from
radiation pressure fluctuations in a thermal state. In order to make
the comparison with standard models of quantum Langevin equations
explicit, a first part briefly recalls the properties related with
fluctuation- dissipation relations and with instability, in the case
of a linear coupling between particle and field. A second part
describes the fluctuating radiation pressure and the radiative force
exerted on a particle scattering a scalar field in a (2d) space-time.
The dependence of the motional susceptibility on the field state is
explicited in the case of thermal input states.
Neglecting recoil effects in a consistent way is shown to lead to
stable and causal motions.

\section{LINEAR COUPLING}

In this part, we recall some general properties of quantum Langevin
equations resulting from linear coupling between a small system and a
bath of oscillators. Emphasis will be put on the properties  of the
motional susceptibility $\chi$ and their consequences for the motions
of the small system.

\subsection{Linear response relations}

Langevin equations are obtained by
coupling a small system, with a few degrees of freedom, to a bath
composed of a very large (infinite) number of degrees of freedom
\cite{lq}.
Eliminating the bath's variables in the equations of motion provides a reduced
equation which involves, besides
the small system's degrees of freedom, noise variables describing
the initial (fluctuating) values of the bath's variables.
In simple examples of quantum Langevin equations, the small system
is linearly coupled to an infinity of harmonic oscillators
representing the bath's degrees of freedom. A standard model is
provided by a particle whose position $q$ (or velocity)
is linearly coupled to a scalar
field in a (2d) space-time $\phi(t,x)$
(representing one polarisation
of the electromagnetic field in a transmission line, for instance).
For a harmonically bound non relativistic particle the Lagrangian can
be written (units will be used such that light velocity is
equal to $1$; the non relativistic limit will be included in linear
response $\dq \ll 1$):
\beqa
\label{ll}
{\it L} = {m\over2} \dq^2 - {K\over 2} q^2 &-& \int_{-\infty}^\infty dx
[a(x) \dq \partial_t \phi + b(x) \dq \phi +
d(x) q \phi] \nonumber\\
&+& \int_{-\infty}^\infty dx
{1\over2} [\partial_t \phi^2 - \partial_x \phi^2]
\eeqa
(a dot meaning time differentiation, and noting that terms like
$q \partial_t \phi$ are
equivalent to terms like $\dq \phi$).
It will also be convenient to use Fourier tranforms in space and time
variables, which will be generally denoted:
\beqa
f(t) = \int_{-\infty}^\infty {d\omega \over 2\pi}  f[\omega] e^{-i\omega t}
\nonumber\\
f(x) = \int_{-\infty}^\infty {dk \over 2\pi} f[k]  e^{ikx}\nonumber
\eeqa
Lagrangian (\ref{ll}) leads to the following solution for the field:
$$\phi[\omega,k] = \phi^{in}[\omega,k] - \chi_{\phi \phi}[\omega,k] e[\omega,k]
q[\omega]$$
with shorthand notation for coupling:
$e[\omega,k] = \omega^2 a[k] - i\omega b[k]
+ d[k] $ (for real coupling $e[\omega,k]^* = e[-\omega,-k]$),
where $\chi_{\phi \phi}$ is the retarded propagator of the field:
\beqn
\label{p1}
\chi_{\phi \phi}[\omega,k] = {-1\over (\omega +
i\epsilon)^2 - k^2}
\eeqn
and $\phi^{in}$ is a free input field.
Using the canonical commutation relations, the retarded propagator can
also be deduced from the free field commutator:
\beqa
\label{p2}
[\phi^{in}(t,x), \phi^{in}(t',x')] &=& 2\hbar \xi_{\phi \phi}(t-t',x-x')\\
\chi_{\phi \phi}(t,x) &=& 2i\theta (t) \xi_{\phi \phi} (t,x)\nonumber
\eeqa
with:
\beqa
\label{xi}
\xi_{\phi \phi}[\omega,k] &=& {\pi\over 2k}[\delta(k-\omega) -
\delta(k+\omega)]\nonumber\\&=& Im(\chi_{\phi \phi}[\omega,k])
\eeqa
These identities relate the field susceptibility to an applied
source
(retarded propagator) with the field spectral density (field
commutator) and  are characteristic of linear response theory \cite{Kubo}.

A quantum Langevin equation follows for the particle:
\beqn
\label{qle}
m\ddq(t) - \int_{-\infty}^\infty dt' \chi (t-t') q(t') = - K q(t)
 + F(t)
\eeqn
where $\chi$ (shortened notation for $\chi_{FF}$) and $F$ describe the motional
susceptibility and the fluctuating force generated
by coupling to the field:
\beqn
\label{chi}
\chi[\omega] = - \int_{-\infty}^\infty {dk \over 2\pi}\quad {e[\omega,k]
e[-\omega,-k]\over
(\omega +
i\epsilon)^2 - k^2 }
\eeqn
$$F[\omega] = - \int_{-\infty}^\infty {dk\over 2\pi} \quad e[-\omega,-k]
\phi^{in}[\omega,k]$$
The force commutator is deduced from the free field commutator
(\ref{p2}):
\beqa
\label{xif}
[F(t), F(t')] &=& 2\hbar
\xi_{FF}(t-t') \nonumber\\
\xi_{FF}[\omega] &=& {1\over
4\omega}(e[\omega,\omega]e[-\omega,-\omega] +
e[\omega,-\omega]e[-\omega,\omega])
\eeqa
Resulting from (\ref{xi}), a fluctuation-dissipation
relation is also satisfied by the motional susceptibility and the force
commutator
(see (\ref{chi}) and (\ref{xif})):
\beqn
\label{fd1}
\xi_{FF}[\omega] = Im(\chi[\omega])
\eeqn
The imaginary part of the susceptibility is related to the amount of
dissipated energy in a stationary regime \cite{Landau}.

The motional force depends on the particle's position in a causal way
(see (\ref{p1}) and (\ref{p2})):
$\chi (t)$ vanishes for negative values of $t$, or else,
$\chi[\omega]$
is analytic in the upper half complex plane ($Im(\omega)>0$).
The causal properties of the motional susceptibility allow one to
write a
dispersion relation. When $\chi$ decreases sufficiently at
infinity, this relation takes a simple form:
\beqa
\label{dr}
\chi[\omega] &=&  \int_{-\infty}^\infty {d\omega' \over \pi}
{\xi_{FF}[\omega'] \over \omega' -\omega - i\epsilon} \nonumber \\
\chi(t) &=& 2i \theta (t) \xi_{FF} (t)\nonumber
\eeqa
These properties can be used to determine the motional susceptibility from the
force commutator.
For linear coupling between the particle's position and
the field, the motional susceptibility is determined
by the field commutator, and
does not depend on the field state. The classical and quantum Langevin
equations then only differ by their noise \cite{lq}.

Moreover, from
the linear dependence of the force on the input field, there results
that for a
thermal input state the quantum noise is gaussian, like the
classical one. The only difference then lies in the correlation
function:
\beqn
\label{ff}
< F(t)F(t')> - <F(t)>^2 = C_{FF}(t-t')
\eeqn
(for a stationary input state). In quantum case, the force
has a commutator satisfying a fluctuation-dissipation relation
\cite{Kubo} (units are taken such that $k_B = 1$):
\beqn
\label{fd2}
2\hbar \xi_{FF}[\omega] = C_{FF}[\omega] - C_{FF}[-\omega] = (1 - e^{-{\hbar
\omega \over
T}}) C_{FF}[\omega]
\eeqn
At zero temperature, only positive frequency modes contribute to the
noise, as expected for the ground state:
$$C_{FF}[\omega] = 2\hbar \theta[\omega] \xi_{FF}[\omega]$$
In the limit of high temperature ($\hbar \omega \ll T$) the classical
fluctuation- dissipation relation is recovered from (\ref{fd1}) and
(\ref{fd2}):
\beqn
\label{fdc}
Im(\chi[\omega]) = {\omega \over 2T} C_{FF}[\omega]
\eeqn
Usual Brownian motion corresponds to a force with white noise correlations
($D$ is the momentum's diffusion coefficient):
$$C_{FF}[\omega] = 2D$$
There results a damping force proportional to velocity, with a
friction coefficient related to the diffusion coefficient
 \cite{E}:
$$\chi[\omega] = i \xi_{FF}[\omega] = {D \over T} i\omega$$

\subsection{Point charge}

In three dimensional space, a similar situation to (\ref{ll}) is
provided by a point charge $e$ located at position\footnote{bold
face letters denote vectors in three dimensional space} ${\bf q}$,
harmonically bound and coupled to the electromagnetic potential
${\bf A}(t,{\bf x})$. In the Coulomb gauge (${\bf \nabla} {\bf A} = 0$) and in
the dipole approximation,
the Lagrangian reads \cite{Dekker}:
$${\it L} = {m\over2} \dot{{\bf q}}^2 - {K\over 2} {\bf q}^2 - e{\bf
A}({\bf 0})\dot{{\bf q}}
+ {1\over 8\pi} \int_{-\infty}^\infty d{\bf x} [ \partial_t{\bf A}^2 -
({\bf \nabla} \wedge {\bf A})^2]$$
Recoil effects are neglected, so that the system's position is
considered to be linearly coupled to the electromagnetic potential,
evaluated at the mean position of the charge (${\bf A}({\bf 0}) =
{\bf A}(t,{\bf 0})$).

Eliminating the electromagnetic field in the resulting equations of
motion, the point charge  obeys a quantum Langevin equation similar to
(\ref{qle}):
\beqn
\label{qqle}
m\ddot{\bf q}(t) - \int_{-\infty}^\infty dt' \chi (t-t') {\bf q}(t') = -
K {\bf q}(t)
 + {\bf F}(t)
\eeqn
with:
\beqa
\chi[\omega] &=& - {4\over3}\omega^2 \int_{0}^\infty  {k^2 dk \over
\pi}
{e^2\over (\omega + i\epsilon)^2 - k^2} \nonumber\\
{\bf F}[\omega] &=& e \dot{{\bf A}}^{in}({\bf 0})[\omega]\nonumber
\eeqa
where ${\bf A}^{in}$ is a free input electromagnetic
potential.
The integral appearing in $\chi$ is divergent (the charge
self-energy is infinite) and must be renormalised.
A simple regulation is obtained by introducing a form factor $\Omega[k]$
which
decouples the charge from the field modes whose frequency exceeds some
large
cut-off frequency $\Omega$ (\cite{Dekker,FLO}):
\beqa
\chi[\omega] &=& - {4\over3}\omega^2 \int_{0}^\infty  {k^2 dk \over
\pi}
{e^2 \Omega[k] \over (\omega + i\epsilon)^2 - k^2} \nonumber\\
&=& {1\over2} \chi^{''}[0]  \omega^2 - {4\over3}\omega^4
\int_{0}^\infty  {dk \over
\pi}
{e^2 \Omega[k] \over (\omega + i\epsilon)^2 - k^2}\nonumber\\
&=& {1\over2} \chi^{''}[0]\omega^2 +{2\over3}ie^2 \omega^3 + O({1\over
\Omega})\nonumber
\eeqa
with:
$${1\over2} \chi^{''}[0] = {4\over3} e^2 \int_{0}^\infty {dk \over \pi}
\Omega[k]$$
A model of regulator is for instance:
$$\Omega[k] = ({\Omega^2 \over \Omega^2 + k^2})^2$$
$$\chi[\omega] = - {e^2 \over 3}{\Omega^3 \omega^2 \over (\omega +
i\Omega)^2} \qquad \qquad \qquad {1\over2}\chi^{''}[0] = {e^2\over3}
\Omega$$
The susceptibility tends to a constant at infinite frequency. Written
in the frequency domain, the left-hand side of Langevin equation
(\ref{qqle}), behaves like $-m\omega^2{\bf q}[\omega]$ at high frequencies, so
that
the bare mass $m$ can be considered as a high frequency mass. At low
frequencies, a further contribution comes from the field reaction,
which induces a mass correction $\mu$ and
leads to a different quasistatic mass $M$:
\beqn
\label{mc}
M = m + \mu \qquad \qquad \mu = {1\over2} \chi^{''}[0]
\eeqn
In the infinite cut-off limit, the renormalised mass $M$ of the
charge remains finite while the induced mass $\mu$ becomes infinite, so
that the bare mass must be infinitely negative \cite{Dekker,FLO}. This limit
provides a
Langevin equation where the motional force is the well-known
radiative reaction force:
\beqn
\label{al}
M\ddot{{\bf q}} - {2\over3} e^2 \stackrel{...}{\bf q} =
- K {\bf q} + {\bf
F}
\eeqn
The left-hand side is the Abraham-Lorentz equation of
classical electron theory \cite{Rohrlich}.

Recalling the commutators of the free electromagnetic field ($k=|{\bf
k}|$):
\beqa
[{\bf A}_i^{in}(t,{\bf x}), {\bf A}_j^{in}(t',{\bf x'})] &=& 2\hbar \xi_{{\bf
A}_i {\bf
A}_j}(t-t',{\bf x}-{\bf x'})\nonumber\\
\xi_{{\bf A}_i {\bf
A}_j}[\omega,{\bf k}]&=& 4\pi (\delta_{ij} - {{\bf k}_i{\bf k}_j\over k^2})
\xi[\omega,k] \nonumber
\eeqa
one can compute the susceptibility from the force fluctuations:
\beqa
\label{fff}
[{\bf F}_i(t),{\bf F}_j(t')] &=& 2\hbar \delta_{ij} \xi_{{\bf F} {\bf
F}}(t-t') \nonumber\\
\xi_{{\bf F}{\bf F}}[\omega] &=& {4\over 3}\omega^2 \int_0^\infty {k^2dk \over
\pi}
e^2 \Omega[k] \xi[\omega,k] \nonumber\\
&=& {2\over3}e^2 \omega^3 \Omega[\omega]\nonumber
\eeqa
Using  analyticity properties and the fluctuation-dissipation
relation (\ref{fd1}), one can recover the motional
susceptibility from the electromagnetic field fluctuations.
The regularised susceptibility tending to a positive constant at
infinite frequency, the dispersion relation must be written with
at least one subtraction \cite{Nussenzveig}. The static susceptibility
$\chi[0]$ vanishes by translation invariance, so that one can write:
\beqa
\label{dr2}
\chi[\omega] &=&  \omega
 \int_{-\infty}^\infty {d\omega' \over \pi}
{\xi_{{\bf FF}}[\omega'] \over \omega' (\omega' -\omega -
i\epsilon)}\nonumber\\
&=& -2\omega^2 \int_0^\infty {dk \over \pi k}
{\xi_{{\bf FF}}[k] \over  (\omega + i\epsilon)^2 -k^2}
\eeqa
(as $\xi_{{\bf FF}}$ is an odd function of the frequency). The induced
mass
depends on the regulator and diverges in the infinite cut-off limit:
\beqa
\label{mc2}
\mu &=& 2\int_0^\infty {dk \over \pi}
{\xi_{{\bf FF}}[k] \over k^3}
\eeqa
The bare mass must contain an (infinitely) negative counterterm (see \ref{mc}),
 which leaves the quasistatic mass $M$ undetermined.

\subsection{Positivity and instability}

The linear equations of motion for the small system (\ref{qqle}) are
easily solved in the frequency domain. Introducing the mechanical
impedance $Z$ and admittance $Y$ of the system (the notation
$f\lbrace p \rbrace = f[ip]$ relates Laplace
with Fourier transforms), one obtains from (\ref{dr2}):
\beqa
\label{imp}
Z\lbrace p\rbrace = mp + {K\over p} - {\chi\lbrace p \rbrace \over p}
= Y\lbrace p\rbrace ^{-1}\nonumber\\
-{{\chi\lbrace p \rbrace} \over p} = \int_{0}^\infty
{dk \over \pi} {\xi_{\bf FF}[k] \over k}{2p\over p^2 +k^2}
\eeqa
with: $\xi_{{\bf FF}}[k] /k \ge 0$.

The system's velocity is determined in terms of the applied force by:
\beqa
\label{rep}
\dot{{\bf q}}[\omega] = Y[\omega] {\bf F}[\omega]
\eeqa
According to causality, the motional force is a retarded function of
the system's displacement. The susceptibility $\chi$ and the mechanical
impedance $Z$, as  functions of
the frequency, are analytic in the upper half plane $Im(\omega)>0$.
No poles are present in the upper half plane, which
could produce unbounded forces from a finite displacement of the
system. Because of its `closed loop gain' like expression,
the admittance $Y$ requires a closer examination. It is well known
that the
Abraham-Lorentz equation (\ref{al}) possesses `runaway solutions'
leading to unstable motions \cite{Rohrlich}.

The spectral decomposition (\ref{imp}) shows that the motional force for
a point charge is related to the system's velocity through a
positive function \cite{Meixner}. $-\chi / p$ is holomorphic in the
complex half plane $Re(p)>0$ and satisfies:
$$Re(-{\chi \lbrace p \rbrace \over p}) > 0 \qquad for \qquad Re(p)>0$$
According to (\ref{mc}, \ref{mc2}), the system's bare mass  remains
positive as long as the cut-off satisfies the inequality:
\beqn
\label{pc}
m \ge 0 \qquad \qquad or \qquad \qquad M \ge \mu
\eeqn
so that the system's impedance is also a positive function in this case.
The inverse of a positive function is positive, and causality
follows from positivity  \cite{Meixner}. When
inequality (\ref{pc}) is satisfied, the admittance is also a causal function,
and no `runaway solutions' can appear (see (\ref{rep})).
The quantum Langevin equation leads to stable motions in this
case \cite{Dekker}.

However, the renormalised impedance and admittance of the point
charge are not positive functions, a consequence of the occurence of
a negative coefficient ($m$) in the spectral
decompositions. Indeed, the renormalised expressions can be written:
$$Z[\omega] = -iM\omega + i{K\over \omega} + {2\over3}e^2\omega^2
= Y[\omega] ^{-1}$$
showing that the admittance has a pole in the upper half plane at
$\omega \sim i3M/2e^2$ (for $\hbar(K/M)^{1\over2} \ll M$). The
renormalised Langevin equation (\ref{al}) leads to unstable
self-accelerating motions of the system (`runaway solutions').
 If specific boundary
conditions are imposed to exclude these unphysical solutions, the
system's motions can then be shown to anticipate on the applied force
and to violate causality \cite{Rohrlich}. Positivity of the bare
mass (\ref{pc}) is the condition for the Langevin equation to lead to
stable and causal motions \cite{Dekker}.

\section{RADIATION PRESSURE }

A neutral  system which scatters a quantum field experiences a radiation
pressure which vanishes in the average, but still fluctuates
\cite{Barton}. In particular, the radiation pressure is
responsible for the Casimir forces between two bodies
\cite{BM,JR2}. The fluctuating radiation pressure produces long
term cumulative effects: a moving scatterer also experiences a mean force
depending on its motion. Using quantum field theory, the
motional force exerted on a mirror in the vacuum of a scalar field has
been obtained \cite{FD}. For small motions, linear response theory
\cite{Kubo} shows that the motional force is connected with the
fluctuations of the radiation pressure at rest \cite{JR3}. A point
scatterer then obeys a Langevin equation of the form (\ref{le}).
We now study the case of a point system scattering a scalar field in a
(2d) space-time.

\subsection{Point scatterer}

In two dimensional space-time, a point scatterer located at a
position $q$ separates space into two regions. In each of them the
scalar field evolves freely and is the sum of two counterpropagating
components:
\beqa
\Phi(t,x) &=& \phi_{in}(t-x) + \psi_{out}(t+x) \qquad \qquad for \quad
x<q
\nonumber\\
\Phi(t,x) &=& \phi_{out}(t-x) + \psi_{in}(t+x) \qquad \qquad for
\quad x>q \nonumber
\eeqa
The outcoming fields are related to the incoming ones by a scattering
matrix. Neglecting recoil effects, the S-matrix on the scatterer at
rest will be written:
\beqa
\label{S}
\phi_{out}[\omega] &=& s[\omega] \phi_{in}[\omega] + e^{2i\omega
q}r[\omega]\psi_{in}[\omega]\nonumber\\
\psi_{out}[\omega] &=& e^{-2i\omega q}r[\omega]\phi_{in}[\omega] +
s[\omega] \psi_{in}[\omega]
\eeqa
$r$ and $s$ are the reflection and transmission coefficients defined
for the
scatterer at rest at $q = 0$.
 Besides the reality, causality and unitarity properties of
the S-matrix:
\beqa
r^*[\omega] = r[-\omega] \qquad &\qquad& s^*[\omega] =
s[-\omega]\nonumber\\
r, s \qquad analytic \quad &for& \quad Im(\omega)>0\nonumber\\
|r|^2 &+& |s|^2 = 1 \nonumber
\eeqa
a transparency condition will be assumed,  with a cut-off frequency
corresponding to an energy smaller than the mass ($M_0$)
of the scatterer:
\beqa
\label{tr}
r[\omega] \sim 0 \qquad \qquad for\quad
\omega \gg \omega_c\nonumber\\
\hbar \omega_c \ll M_0
\eeqa
This condition
(satisfied by realistic mirrors) allows one to neglect recoil effects,
and will play an important role in the following. A perfect reflector
corresponds to $r=-1$ for all frequencies and does not obey the
required conditions.

In each region, the energy ($e$) and momentum ($p$) densities of the
field
are those of a free scalar field:
\beqa
e(t,x) &=& \dphi^2(t-x) + \dpsi^2(t+x)
\nonumber\\
p(t,x) &=& \dphi^2(t-x) - \dpsi^2(t+x)\nonumber
\eeqa
The radiation pressure exerted on the motionless scatterer is
obtained from the stress tensor of the field, evaluated at the
scatterer's position:
\beqn
\label{F}
F(t) = \dphi_{in}^2(t-q) + \dpsi_{out}^2(t+q) - \dphi_{out}^2(t-q) -
\dpsi_{in}^2(t+q)
\eeqn
and can be expressed in terms of the input fields and the S-matrix.

\subsection{Radiation pressure fluctuations and radiative reaction}

For input fields in a stationary and isotropic state, the field correlation
functions can be written:
\beqa
-\omega\omega'<\phi_{in}[\omega]\phi_{in}[\omega']> &=&
-\omega\omega'<\psi_{in}[\omega]\psi_{in}[\omega']> = 2\pi c[\omega]
\delta(\omega+\omega')\nonumber\\
<\phi_{in}[\omega]\psi_{in}[\omega']> &=& 0 \nonumber
\eeqa
The field correlations can be decomposed into an antisymmetric part
(free field commutator) which does not depend on the state
 and a symmetric part (mean field anticommutator) which
is state dependent (see eq.(\ref{p2}); $\xi$ is a shortened notation
for $\xi_{\dot{\phi}\dot{\phi}}$):
\beqa
-\omega\omega'[\phi_{in}[\omega],\phi_{in}[\omega']] &=&
-\omega\omega'[\psi_{in}[\omega],\psi_{in}[\omega']] = 4\pi\hbar
\xi[\omega]\delta(\omega+\omega')  \nonumber\\
c[\omega] &=& \hbar (\xi[\omega] + \sigma[\omega] ) \qquad
\qquad \xi[\omega] = {\omega \over 4}\nonumber
\eeqa
In particular, for a thermal input state the field correlations are
given by a fluctuation-dissipation relation (see \ref{fd2}):
\beqa
c[\omega] &=& {2\hbar \xi[\omega] \over 1 - e^{-{\hbar \omega
\over T}}} = {\hbar \omega \over 2(1 - e^{-{\hbar \omega
\over T}})}\nonumber\\
\sigma[\omega] &=& {\omega\over4}coth{\hbar\omega \over 2T}\nonumber
\eeqa
The radiation pressure fluctuations are determined by the
fluctuations of the input fields (see eqs (\ref{ff}) and (\ref{F})).
 In a thermal
state, the mean  quartic forms are obtained from the 2-point
correlations using Wick's rules, and lead to (see eq.(25) of
\cite{JR3}; $\omega^2 c[\omega]$ has been changed to $c[\omega]$):
\beqa
\label{C}
C_{FF}[\omega] &=& \int_{-\infty}^\infty {d\omega' \over 2\pi}
4c[\omega']c[\omega-\omega'] \gamma[\omega',\omega-\omega']\nonumber\\
\gamma &=& |\alpha|^2 + |\beta|^2\nonumber\\
\alpha[\omega,\omega'] &=& 1-s[\omega]s[\omega']+r[\omega]r[\omega']
\qquad \nonumber\\
\beta[\omega,\omega'] &=&
s[\omega]r[\omega']-r[\omega]s[\omega']
\eeqa
The mean force commutator follows and satisfies
fluctuation-dissipation relation (\ref{fd2}).
Noting that:
$$\gamma[\omega,\omega']=\gamma[\omega',\omega]=\gamma[-\omega,-\omega']$$
$$(1 - e^{-{\hbar \omega \over T}})
c[\omega']c[\omega-\omega'] = {\hbar^2 \over2} \lbrace
\omega'\sigma[\omega-\omega'] + (\omega-\omega')\sigma[\omega']
\rbrace $$
it can also be written:
$$\xi_{FF}[\omega] = \hbar \int_{-\infty}^\infty {d\omega' \over 2\pi}
2(\omega-\omega')\sigma[\omega']
\gamma[\omega',\omega-\omega']$$
This expression, which also results directly from (\ref{F}) and the field
commutator, exhibits the general dependence of the mean force
commutator on the input state.

Motions of the point scatterer alter the field scattering. The
S-matrix introduced in (\ref{S}) describes the field scattering in
the comoving frame and coordinate transformations must be used to
recover the S-matrix in the original frame \cite{JR3}. The expression
of the force in terms of the scattered fields also suffers velocity
dependent changes following the covariant nature of the field stress
tensor. Considering only first order corrections in
displacements, the radiation pressure exerted on a the moving scatterer
is obtained under the form (\ref{le}) with (see eq.(19) of
\cite{JR3}):
$$\chi[\omega] = 4i\hbar \int_{-\infty}^\infty {d\omega' \over 2\pi}
(\omega-\omega')\sigma[\omega'] \alpha[\omega',\omega-\omega']$$
For a stationary state, the static susceptibility vanishes ($\sigma$
is an even function of $\omega$):
$$\chi[0] = 0$$
Recalling the S-matrix unitarity, one remarks that $\gamma = 2
Re(\alpha)$, so that fluctuation-dissipation (\ref{fd1}) is satisfied
by the radiation pressure fluctuations and the radiative reaction
force.

In contrast to linear coupling, the radiative
reaction force exerted on a point scatterer depends on the input field
state. The susceptibility for a thermal state at temperature $T$:
$$\chi_T[\omega] = i\hbar \int_{-\infty}^\infty {d\omega' \over 2\pi}
\omega'(\omega-\omega') \lbrace 1 + {2 \over  e^{\hbar \omega' \over
T} - 1} \rbrace \alpha[\omega',\omega-\omega']$$
can be decomposed into a vacuum contribution and a
thermal correction:
$$\chi_0[\omega] = i\hbar \int_0^\omega {d\omega' \over
2\pi} \omega'(\omega-\omega')\alpha[\omega',\omega-\omega']$$
\beqa
\label{chiT}
\chi_T[\omega] = \chi_0[\omega] + 2i\hbar \int_0^\infty {d\omega' \over
2\pi}{\omega' \over e^{\hbar \omega' \over T} -
1}\lbrace(\omega+\omega')\alpha[-\omega',\omega+\omega'] \nonumber\\
+(\omega-\omega')\alpha[\omega',\omega-\omega']\rbrace
\eeqa
The force responses to quasistatic motions (translation, constant
velocity, constant acceleration, ...) are given by a Taylor expansion
around zero frequency:
$$ \chi_0'[0] = \chi_0^{''}[0] = 0$$
The vacuum state is  Lorentz invariant. Under uniformly
accelerated motion its  fluctuations appear as thermal ones
in the
comoving frame of the point scatterer \cite{Boyer}. Hence, the corresponding
responses vanish. At non zero temperature, a friction coefficient
related to viscosity
and a correction to the quasistatic mass appear:
\beqa
\label{chi0}
\chi_T'[0] =  2i\hbar  \int_0^\infty {d\omega' \over
2\pi} {\omega' \over e^{\hbar \omega'
\over T} -
1}&\lbrace&(1+\omega'\partial_{\omega'})\alpha[\omega',-\omega']\nonumber\\
&+&(1-\omega'\partial_{\omega'})\alpha[-\omega',\omega'] \rbrace\nonumber\\
{1\over2}\chi_T^{''}[0] = 2i\hbar \int_0^\infty
{d\omega' \over2\pi} {\omega' \over e^{\hbar \omega'
\over T} - 1}&\lbrace&(1+{\omega' \over 2}\partial_{\omega'})
\partial_{\omega'}\alpha[\omega',-\omega']\nonumber\\
&+&(1-{\omega' \over
2}\partial_{\omega'})\partial_{\omega'}\alpha[-\omega',\omega'] \rbrace
\eeqa
The point scatterer obeys a Langevin equation where the
susceptibility and the force fluctuations are given by (\ref{chiT}),
and fluctuation-dissipation relations (\ref{fd1}) and (\ref{fd2}):
\beqn
\label{sle}
M_0\ddq(t) - \int_{-\infty}^\infty dt' \chi_T(t-t') q(t') = -
K q(t)
 + F(t)
\eeqn
The quasistatic responses
(\ref{chi0}) show that the quasistatic mass depends on the
temperature, and that the mass entering the Langevin equation  is
the vacuum quasistatic mass $M_0$:
\beqn
\label{MT}
M_T = M_0 + {1\over2}\chi_T^{''}[0]
\eeqn
Expanding  around zero temperature, the first terms of the susceptibility
$\chi_T$  and of the force commutator
$\xi_T$ can be obtained:
\beqa
\chi_T[\omega] &=& \chi_0[\omega]
+ {i\pi
T^2 \over
3\hbar} \omega \alpha[0,\omega]\nonumber\\
\xi_T[\omega] &=& \xi_0[\omega]
 + {i\pi
T^2 \over
6\hbar} \omega \gamma[0,\omega]\nonumber\\
\xi_0[\omega] &=& i\hbar \int_0^\omega {d\omega' \over
4\pi} \omega'(\omega-\omega')\gamma[\omega',\omega-\omega']\nonumber
\eeqa
Temperature corrections induce a damping force proportional to the
velocity as in Brownian motion. The friction coefficient vanishes
like  $T^2$ near vacuum.

For temperature and frequencies well below the
reflection cut-off ($T \ll \hbar \omega_c , \omega \ll \omega_c$),
and if the scatterer can be considered as a perfect reflector over a large
frequency interval,
 the limit of constant reflectivity
($\alpha[\omega,\omega']
= 2$) can be taken in (\ref{chiT}), resulting in a simple form for
the susceptibility:
\beqa
\chi_T[\omega] &=& i\xi_T[\omega]\nonumber\\
\xi_T[\omega] &=& {\hbar \over 6\pi} \omega^3 + {2\pi T^2
\over
3\hbar} \omega
\eeqa
In the classical limit ($\hbar \omega \ll T$), Brownian motion \cite{E}
is recovered with a diffusion coefficient for the particle's momentum
(see (\ref{fdc})):
\beqn
D = T \xi_T'[0] = {2\pi T^3
\over
3\hbar}
\eeqn
In vacuum,  the momentum diffusion vanishes (a consequence of
momentum conservation) and  a Langevin
equation similar to the Abraham-Lorentz equation for a point charge
follows:
$$M_0\ddot{q} - {\hbar \over 6\pi} \stackrel{...}{q}   = - Kq + F$$
A perfect reflector in vacuum is affected by the same instability
problems as the point charge. Recalling (\ref{tr}), this illustrates
the incompatibility between the infinite cut-off limit and the
approximation neglecting recoil effects. In next section we show how
stability and causality follow from  a consistent treatment of
radiative reaction.

\subsection{Passivity}

 The causal nature of the force susceptibility results from
that of the S-matrix \cite{JR3}, so that analytic properties can be
used to recover the motional susceptibility from the mean force commutator and
a dispersion relation.
{}From eq.(\ref{chiT}), the
high frequency
behavior of the susceptibility is dominated by
the vacuum
contribution. We shall assume in the following
that the
reflectivity is cutted off at high frequencies and
that the vacuum
susceptibility is such that
$\chi_0[\omega] /
\omega^3$
 is a square integrable function.
(A model of S-matrix
satisfying causality, unitarity and transparency
is for instance:
$$s=1+r \qquad r[\omega]=-{i\Omega \over
\omega + i\Omega} \qquad
{\chi_0[\omega] \over
\omega^3 }\sim -{\hbar\Omega \over 2\pi\omega} \quad for \quad
\omega \gg
\Omega)$$
Then, the
dispersion relation in the vacuum state can be written
\cite{Nussenzveig}:
$$\chi_0[\omega] = \omega^3 \int_{-\infty}^\infty
{d\omega' \over \pi \omega'^3}
{\xi_0[\omega'] \over \omega' -\omega -
i\epsilon}$$
so that at high frequencies:
\beqn
\label{mu0}
{\chi_0[\omega] \over \omega^3} \sim -{\hbar\omega_c \over \omega}
\qquad \qquad
\hbar\omega_c = \mu_0 = 2\int_0^\infty {dk \over
\pi}
{\xi_0[k] \over k^3} \quad < \infty
\eeqn
This defines the cut-off frequency $\omega_c$ introduced in
(\ref{tr}). The thermal correction satisfies:
\beqa
\label{muT}
\chi_T[\omega] - \chi_0[\omega] &=& \chi_T'[0]\omega + \omega^2
\int_{-\infty}^\infty
{d\omega' \over \pi \omega'^2} {\xi_T[\omega'] - \xi_0[\omega'] -
\xi_T'[0]\omega'
\over \omega' - \omega -i\epsilon} \nonumber\\
{1\over2}\chi_T^{''}[0] &=& \mu_T - \mu_0 \nonumber\\
\mu_T &=& \int_{-\infty}^\infty
{dk \over \pi} {\xi_T[k] - \xi_T'[0]k \over k^3}
\eeqa
In a thermal state, the susceptibility (\ref{chiT}) is recovered from the force
commutator using the
dispersion relation ($\xi_T^{''}[0]=0$):
$$\chi_T[\omega] = \chi_T'[0]\omega + {1\over2}\chi_T^{''}[0]\omega^2 +
\omega^3 \int_{-\infty}^\infty {d\omega' \over \pi \omega'^3}
{\xi_T[\omega'] - \xi_T'[0]\omega' \over \omega' - \omega -i\epsilon}$$
 or in  Laplace transforms:
\beqa
-{\chi_T\lbrace p \rbrace \over p} &=& -i\chi_T'[0] +
{1\over2}\chi_T^{''}[0]p - p^2\int_{-\infty}^\infty {dk \over \pi k^3}
{\xi_T[k] - \xi_T'[0]k \over p + ik}\nonumber\\
&=& \xi_T'[0]
- \mu_0 p
+ \int_{-\infty}^\infty {dk\over \pi}
{\xi_T[k] \over k(1+k^2)}{1 + ipk \over p + ik}\nonumber
\eeqa
{}From (\ref{C}) the coefficients entering the spectral decomposition
can be seen to satisfy:
$$\xi_T'[0] \ge 0 \qquad \qquad
{\xi_T[k] \over k} \ge 0$$
It results that for a point scatterer $-\chi/p$ is not a positive function
(it has a negative residue for the pole
at infinity).

 Langevin equation for the scatterer (\ref{sle}) is solved in the
frequency domain by (\ref{rep}), where the scatterer's impedance and
admittance are given by:
\beqa
Z\lbrace p\rbrace &=& \xi_T'[0] + (M_0 - \mu_0)p
 + {K\over p} \nonumber\\
&+& \int_{-\infty}^\infty {dk \over \pi
}
{\xi_T[k] \over k(1+k^2)}{1 + ipk  \over p + ik}
= Y\lbrace p\rbrace ^{-1}\nonumber
\eeqa
The high frequency mass $m$ of the scatterer is related to the
quasistatic mass $M_T$ through (see (\ref{MT}) and (\ref{muT})):
$$m = M_T - \mu_T = M_0 - \mu_0 $$
The scatterer's quasistatic mass is greater than its high frequency mass
(see (\ref{mu0})). The
difference
is  a mass induced by the field swept along the
scatterer's motion, and vanishing at high frequencies  where
field and scatterer decouple (see the transparency condition). When
(\ref{tr}) is satisfied,  the high frequency mass is positive and
 the system's impedance has the spectral decomposition of
 a positive (or passive)
function \cite{Meixner}. The admittance is  also a positive function and
the Langevin equation leads to stable and causal motions for
a scatterer in a thermal state near vacuum \cite{JR5}. Positivity
of the high frequency mass, or  a
quasistatic mass greater than the induced mass,
 is the condition for stable and causal
motions of the scatterer. It follows from that description
that perfect reflection can only
be consistent with an infinite quasistatic mass.  For a finite mass
scatterer, recoil effects must be taken into account before
considering the reflection cut-off as infinite.

\section{CONCLUSION}

Quantum Langevin equations possess general properties which are
characteristic of linear response theory \cite{lq,Kubo}. Coupling a
particle to a field through radiation pressure leads to pecular
properties. The radiation pressure fluctuations and the radiative
reaction force satisfy fluctuation- dissipation relations which
exhibit a dependence of Langevin equation's kernel upon the input field
state. The relations have been obtained here for thermal fields
scattered by a point system. A similar situation occurs when
irradiating a mirror with coherent light: the mirror
satisfies a Langevin equation which depends on the intensity of the
incident light. This has consequences on the ultimate sensitivity of
interferometric measurements of positions \cite{Unruh}.

Consistency of the simplified description in terms of reflection
and transmission coefficients requires that the scatterer be
transparent at frequencies greater than a reflection cut-off,
corresponding to an energy
smaller than the scatterer's mass. This inequality implies that the
mass induced by field reaction is smaller than the quasistatic mass,
i.e. that the high frequency mass of the scatterer
is positive. This identifies with the condition for the Langevin
equation to lead to stable and causal motions. This property can be
considered as a consequence of the passivity of states near vacuum,
i.e. of their incapacity to sustain `runaway solutions' \cite{JR5}.
This must be compared with the case of a point charge, linearly
coupled to the electromagnetic field. Renormalisation leads to an
infinite negative bare mass, so that the Langevin equation possesses
unstable solutions or violates causality \cite{Dekker}. In this case, recoil
effects should explicitly be taken into account to get a consistent
treatment of radiative reaction.

\begin{flushleft}
{\bf Acknowledgement}
\end{flushleft}

This paper is dedicated to the memory of R.Rammal. The problems
discussed here belong to some of the many subjects, which R.Rammal took an
interest in and enjoyed  sharing reflections upon.

\end{document}